# Modified Structure of Protons and Neutrons in Correlated Pairs

B. Schmookler, M. Duer, A. Schmidt, O. Hen, S. Gilad, E. Piasetzky, M. Strikman, L.B. Weinstein et al.
(The CLAS Collaboration)

**The atomic nucleus is made of protons and neutrons (nucleons), that are themselves composed of quarks and gluons. Understanding how the quark-gluon structure of a nucleon bound in an atomic nucleus is modified by the surrounding nucleons is an outstanding challenge. Although evidence for such modification, known as the EMC effect, was first observed over 35 years ago, there is still no generally accepted explanation of its cause [1–3]. Recent observations suggest that the EMC effect is related to close-proximity Short Range Correlated (SRC) nucleon pairs in nuclei [4, 5]. Here we report the first simultaneous, high-precision, measurements of the EMC effect and SRC abundances. We show that the EMC data can be explained by a universal modification of the structure of nucleons in neutron-proton (*np*) SRC pairs and present the first data-driven extraction of this universal modification function. This implies that, in heavier nuclei with many more neutrons than protons, each proton is more likely than each neutron to belong to an SRC pair and hence to have its quark structure distorted.**

We study nuclear and nucleon structure by scattering high-energy electrons from nuclear targets. The energy and momentum transferred from the electron to the target determines the space-time resolution of the reaction, and thereby, which objects are probed (i.e., quarks or nucleons). To study the structure of nuclei in terms of individual nucleons, we scatter electrons in quasi-elastic (QE) kinematics where the transferred momentum typically ranges from 1 to 2 GeV/c and the transferred energy is consistent with elastic scattering from a moving nucleon. To study the structure of nucleons in terms of quarks and gluons, we use Deep Inelastic Scattering (DIS) kinematics with larger transferred energies and momenta.

Atomic nuclei are broadly described by the nuclear shell model, in which protons and neutrons move in well-defined quantum orbitals, under the influence of an average mean-field created by their mutual interactions. The internal quark-gluon substructure of nucleons was originally expected to be independent of the nuclear environment because quark interactions occur at shorter-distance and higher-energy scales than nuclear interactions. However, DIS measurements indicate that quark momentum distributions in nucleons are modified when nucleons are bound in atomic nuclei [1, 2, 6, 7], breaking down the scale separation between nucleon structure and nuclear structure.

This scale separation breakdown in nuclei was first observed thirty-five years ago in DIS measurements performed by the European Muon Collaboration (EMC) at CERN [8]. These showed a decrease of the DIS cross-section ratio of iron to deuterium in a kinematical region corresponding to moderate- to high-momentum quarks in the bound nucleons. The EMC effect has been confirmed by subsequent measurements on a wide variety of nuclei, using both muons and electrons [9, 10], and over a large range of transferred momenta, see reviews in [1, 2, 6, 7]. The maximum reduction in the DIS cross-section ratio of a nucleus relative to deuterium increases from about 10% for $^4$He to about 20% for Au.

The EMC effect is now largely accepted as evidence that quark momentum distributions are different in bound nucleons relative to free nucleons [1, 2, 7]. However, there is still no consensus as to the underlying nuclear dynamics driving it.

Currently, there are two leading approaches for describing the EMC effect, which are both consistent with data: (A) all nucleons are slightly modified when bound in nuclei, or (B) nucleons are unmodified most of the time, but are modified significantly when they fluctuate into SRC pairs. See Ref. [1] for a recent review.

SRC pairs are temporal fluctuations of two strongly-interacting nucleons in close proximity, see e.g. [1, 11]. Electron scattering experiments in QE kinematics have shown that SRC pairing shifts nucleons from low-momentum nuclear shell-model states to high-momentum states with momenta greater than the nuclear Fermi momentum. This "high-momentum tail" has a similar shape for all nuclei. The relative abundance of SRC pairs in a nucleus relative to deuterium approximately equals the ratio of their inclusive (e,e′) electron scattering cross-sections in selected QE kinematics [12–15].

Recent studies of nuclei from $^4$He to Pb [16–22], showed that SRC nucleons are "isophobic"; i.e., similar nucleons are much less likely to pair than dissimilar nucleons, leading to many more *np* SRC pairs than neutron-neutron (*nn*) and proton-proton (*pp*) pairs. The probability for a neutron to be part of an *np*-SRC pair is observed to be approximately constant for all nuclei, while that for a proton increases approximately as *N/Z*, the relative number of neutrons to protons [22].

The first experimental evidence supporting the SRC-modification hypothesis as an explanation for the EMC effect came from comparing the abundances of SRC pairs in different nuclei with the size of the EMC effect. Not only do both increase from light to heavy nuclei, but there is a robust linear correlation between them [4, 5]. This suggests that the EMC effect might be related to the high-momentum nucleons in nuclei.

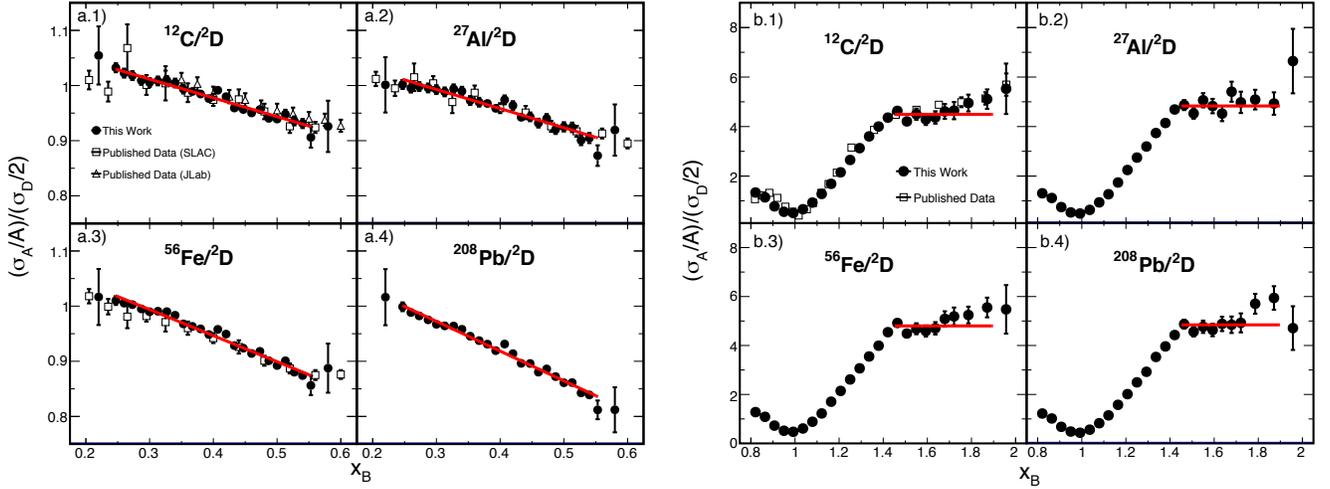

**Fig 1 | DIS and QE (e,e′) Cross-section Ratios.** The per-nucleon cross-section ratios of nucleus with atomic number A to deuterium for (a. 1 - 4) DIS kinematics ($0.2 \leq x_B \leq 0.6$ and $W \geq 1.8$ GeV). The solid points show the data of this work, the open squares the data of [9] and the open triangles show the data of [10]. The red lines show the linear fit. (b. 1 - 4) QE kinematics ($0.8 \leq x_B \leq 1.9$). The solid points show the data of this work and the open squares the data of [11]. The red lines show the constant fit. The error bars shown include both statistical and point-to-point systematic uncertainties, both at the 1σ or 68% confidence level. The data are not isoscalar corrected.

The analysis reported here was motivated by the quest to understand the underlying patterns of nucleon structure modification in nuclei and how this varies from symmetric to asymmetric nuclei. We measured both the DIS and QE inclusive cross-sections simultaneously for deuterium and heavier nuclei, thereby reducing the uncertainties in the extraction of the EMC effect and SRC scaling factors. We observed that: (1) the EMC effect in all measured nuclei is consistent with being due to the universal modification of the internal structure of nucleons in np-SRC pairs, permitting the first data-driven extraction of this universal modification function, (2) the measured per-proton EMC effect and SRC probabilities continue to increase with atomic mass A for all measured nuclei while the per-neutron ones stop increasing at $A \approx 12$, and (3) the EMC-SRC correlation is no longer linear when the EMC data are not corrected for unequal numbers of proton and neutrons. We also constrained the internal structure of the free neutron using the extracted universal modification function and we concluded that in neutron-rich nuclei the average proton structure modification will be larger than that of the average neutron.

We analyzed experimental data taken using the CLAS spectrometer [23] at the Thomas Jefferson National Accelerator Facility (Jefferson Lab). In our experiment, a 5.01 GeV electron beam impinged upon a dual target system with a liquid deuterium target cell followed by a foil of either C, Al, Fe or Pb [24]. The scattered electrons were detected in CLAS over a wide range of angles and energies which allowed extracting both QE and DIS reaction cross-section ratios over a wide kinematical region (See Supplementary Information section I).

The electron scattered from the target by exchanging a single virtual photon with momentum $\vec{q}$ and energy $\nu$, giving a four-momentum transfer $Q^2 = |\vec{q}|^2 - \nu^2$. We used these variables to calculate the invariant mass of the nucleon plus virtual photon $W^2 = (m + \nu)^2 - |\vec{q}|^2$ (where m is the nucleon mass) and the scaling variable $x_B = Q^2/2m\nu$.

We extracted cross-section ratios from the measured event yields by correcting for experimental conditions, acceptance and momentum reconstruction effects, reaction effects, and bin-centering effects. See Supplementary Information section I. This was the first precision measurement of inclusive QE scattering for SRCs in both Al and Pb, as well as the first measurement of the EMC effect on Pb. For other measured nuclei our data are consistent with previous measurements but with reduced uncertainties.

The DIS cross-section on a nucleon can be expressed as a function of a single structure function, $F_2(x_B, Q^2)$. In the parton model, $x_B$ represents the fraction of the nucleon momentum carried by the struck quark. $F_2(x_B, Q^2)$ describes the momentum distribution of the quarks in the nucleon, and the ratio, $[F_2^A(x_B, Q^2)/A] / [F_2^d(x_B, Q^2)/2]$, describes the relative quark momentum distributions in nucleus A and deuterium [2, 7]. For brevity, we will often omit explicit reference to $x_B$ and $Q^2$, i.e., writing $F_2^A/F_2^d$, with the understanding that the structure functions are being compared at identical $x_B$ and $Q^2$. Because the DIS cross-section is proportional to $F_2$, experimentally the cross-section ratio of two nuclei is assumed to equal their structure-function ratio [1, 2, 6, 7]. The magnitude of the EMC effect is defined by the slope of either the cross-section or the structure-function ratios for $0.3 \leq x_B \leq 0.7$ (see Supplementary Information sections IV and V).

Similarly, the relative probability for a nucleon to belong to an SRC pair is interpreted as equal to $a_2$, the average value of the inclusive QE electron-scattering per-nucleon cross-section ratios of nucleus A compared to deuterium at momentum transfer $Q^2 > 1.5$ GeV$^2$ and $1.45 \leq x_B \leq 1.9$ [1, 11-15] (see Supplementary Information section III). Other nuclear effects are expected to be negligible. The contribution of three-nucleon SRCs should be an order of magnitude smaller than the SRC pair contributions. The contributions of two-body currents (called "higher-twist effects" in DIS scattering) should also be small (see Supplementary Information section VIII).

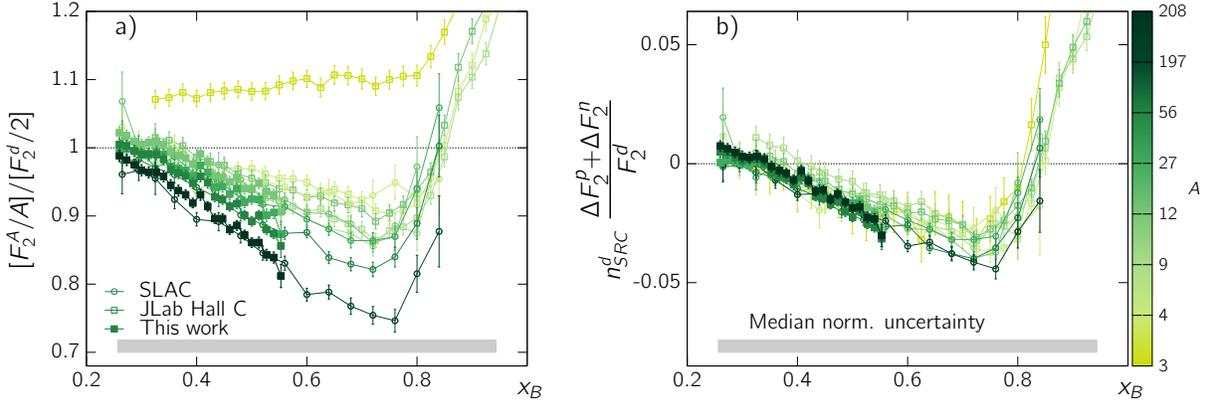

**Fig 2 | Universality of SRC pair quark distributions.** The EMC effect for different nuclei, as observed in (a) ratios of $(F_2^A/A)/(F_2^d/2)$ as a function of $x_B$ and (b) the modification of SRC pairs, as described by Eq. 2. Different colors correspond to different nuclei, as indicated by the color scale on the right. The open circles show SLAC data [9] and the open squares show Jefferson Lab data [10]. The nucleus-independent (universal) behavior of the SRC modification, as predicted by the SRC-driven EMC model, is clearly observed. The error bars on the symbols show both statistical and point-to-point systematic uncertainties, both at the 1σ or 68% confidence level and the gray bands show the median normalization uncertainty. The data are not isoscalar corrected.

Figure 1 shows the DIS and QE cross-section ratios for scattering off the solid target relative to deuterium as a function of $x_B$. The red lines are fits to the data that are used to determine the EMC effect slopes or SRC scaling coefficients (see Extended Data Table I and II). Typical 1σ cross-section ratio normalization uncertainties of 1 − 2% directly contribute to the uncertainty in the SRC scaling coefficients but introduce a negligible EMC slope uncertainty. None of the ratios presented have isoscalar corrections (cross-section corrections for unequal numbers of protons and neutrons), in contrast to much published data. We do this for two reasons, (1) to focus on asymmetric nuclei and (2) because the isoscalar corrections are model-dependent and differ among experiments [9, 10] (see Extended Data Fig. 1).

The DIS data was cut on $Q^2 > 1.5$ GeV$^2$ and $W > 1.8$ GeV, which is just above the resonance region [25] and higher than the $W > 1.4$ GeV cut used in previous JLab measurements [10]. The extracted EMC slopes are insensitive to variations in these cuts over $Q^2$ and $W$ ranges of $1.5 − 2.5$ GeV$^2$ and $1.8 − 2$ GeV respectively (see Supplementary Information Table VII).

Motivated by the correlation between the size of the EMC effect and the SRC pair density ($a_2$), we model the modification of the nuclear structure function, $F_2^A$, as due entirely to the modification of np-SRC pairs. $F_2^A$ is therefore decomposed into contributions from unmodified mean-field protons and neutrons (the first and second terms in Eq. 1), and np-SRC pairs with modified structure functions (third term):

$$F_2^A = (Z - n_{SRC}^A)F_2^p + (N - n_{SRC}^A)F_2^n + n_{SRC}^A(F_2^{p*} + F_2^{n*}) \quad \text{Eq. 1}$$
$$= ZF_2^p + NF_2^n + n_{SRC}^A(\Delta F_2^p + \Delta F_2^n),$$

where $n_{SRC}^A$ is the number of np-SRC pairs in nucleus $A$, $F_2^p(x_B, Q^2)$ and $F_2^n(x_B, Q^2)$ are the free proton and neutron structure functions, $F_2^{p*}(x_B, Q^2)$ and $F_2^{n*}(x_B, Q^2)$ are the average modified structure functions for protons and neutrons in SRC pairs, and $\Delta F_2^n = F_2^{n*} - F_2^n$ (and similarly for $\Delta F_2^p$). $F_2^{p*}$ and $F_2^{n*}$ are assumed to be the same for all nuclei. In this simple model, nucleon motion effects [1–3], which are also dominated by SRC pairs due to their high relative momentum, are folded into $\Delta F_2^p$ and $\Delta F_2^n$.

This model resembles that used in [26]. However, that work focused on light nuclei and did not determine the shape of the modification function. Similar ideas using factorization were discussed in [1], such as a model-dependent ansatz for the modified structure functions which was shown to be able to describe the EMC data [27]. The analysis presented here is the first data-driven determination of the modified structure functions for nuclei from $^3$He to lead.

Since there are no model-independent measurements of $F_2^n$, we apply Eq. 1 to the deuteron, rewriting $F_2^n$ as $F_2^d - F_2^p - n_{SRC}^d(\Delta F_2^p + \Delta F_2^n)$. We then rearrange Eq. 1 to get:

$$\frac{n_{SRC}^d(\Delta F_2^p + \Delta F_2^n)}{F_2^d}$$
$$= \frac{\frac{F_2^A}{F_2^d} - (Z - N)\frac{F_2^p}{F_2^d} - N}{(A/2)a_2 - N}, \quad \text{Eq. 2}$$

where $F_2^p/F_2^d$ was previously measured [28] and $a_2$ is the measured per-nucleon cross-section ratio shown by the red lines in Fig. 1b. Here we assume $a_2$ approximately equals the per-nucleon SRC-pair density ratio of nucleus $A$ and deuterium: $(n_{SRC}^A/A)/(n_{SRC}^d/2)$ [1, 11-15].

Since $\Delta F_2^p + \Delta F_2^n$ is assumed to be nucleus-independent, our model predicts that the left-hand side of Eq. 2 should be a universal function (i.e., the same for all nuclei). This requires that the nucleus-dependent quantities on the right-hand side of Eq. 2 combine to give a nucleus-independent result.

This is tested in Fig. 2. The left panel shows $[F_2^A(x_B)/A] / [F_2^d(x_B)/2]$, the per-nucleon structure-function ratio of different nuclei relative to deuterium without isoscalar corrections. The approximately linear deviation from unity for $0.3 \leq x_B \leq 0.7$ is the EMC effect, which is larger for heavier nuclei. The right panel shows the relative structure modification of nucleons in $np$-SRC pairs, $n_{SRC}^d (\Delta F_2^p + \Delta F_2^n)/F_2^d$, extracted using the right-hand side of Eq. 2.

The EMC slope for all measured nuclei increases monotonically with $A$ while the slope of the SRC-modified structure function is constant within uncertainties, see Fig. 3 and Extended Data Table II. Even $^3$He, which has a dramatically different structure-function ratio due to its extreme proton-to-neutron ratio of 2, has a remarkably similar modified structure function with the same slope as the other nuclei. Thus, we conclude that the magnitude of the EMC effect in different nuclei can be described by the abundance of $np$-SRC pairs and that the proposed SRC-pair modification function is, in fact, universal. This universality appears to hold even beyond $x_B = 0.7$.

The universal function extracted here will be tested directly in the future using lattice QCD calculations [26] and by measuring semi-inclusive DIS off the deuteron, tagged by the detection of a high-momentum backward-recoiling proton or neutron that will allow to directly quantify the relationship between the momentum and the structure-function modification of bound nucleons [29].

The universal SRC-pair modification function can also be used to extract the free neutron-to-proton structure-function ratio, $F_2^n/F_2^p$, by applying Eq. 1 to the deuteron and using the measured proton and deuteron structure functions (see Extended Data Fig. 1). In addition to its own importance, this $F_2^n$ can be used to apply self-consistent isoscalar corrections to the EMC effect data (see Supplementary Information Eq. 5).

To further test the SRC-driven EMC model, we consider the isophobic nature of SRC pairs (i.e., $np$-dominance), which leads to an approximately constant probability for a neutron to belong to an SRC pair in medium to heavy nuclei, while the proton probability increases as $N/Z$ [22]. If the EMC effect is indeed driven by high-momentum SRCs, then in neutron-rich nuclei both the neutron EMC effect and the SRC probability should saturate, while for protons both should grow with the nuclear mass and the neutron excess.

This is done by examining the correlation of the individual per-proton and per-neutron QE SRC cross-section ratios, $a_2^p = (\sigma_A/Z)/\sigma_d$ and $a_2^n = (\sigma_A/N)/\sigma_d$, and DIS EMC slopes, $dR_{EMC}^p/dx_B$ and $dR_{EMC}^n/dx_B$ (see Extended Data Tables I and III and Supplementary Information sections III and V).

Figure 4 shows the per-proton and per-neutron EMC slopes as a function of $a_2^p$ and $a_2^n$, respectively. We consider these correlations both before (top panels) and after (bottom panels) applying isoscalar corrections to the EMC data and compare them with the predictions of the SRC-driven EMC model. By not applying isoscalar corrections, the top panel allows focusing on the separate behavior of protons and neutrons. Applying self-consistent isoscalar corrections makes both the per-neutron and per-proton EMC-SRC correlations linear, in overall agreement with the model prediction for $N = Z$ nuclei.

This simple rescaling of the previous EMC-SRC correlation result [4, 5], as expected, does not change the EMC-SRC correlation or its slope. However, the per-neutron and per-proton results differ significantly. Because the probability that a neutron belongs to an SRC pair does not increase for nuclei heavier than C ($A = 12$) [22], our model predicts that the per-neutron EMC effect (i.e., the slope of $\frac{F_2^A/N}{F_2^d/1}$) will also not increase for $A \geq 12$.

In contrast, the probability that a proton belongs to an SRC pair continues to increase for all measured nuclei [22] and therefore the per-proton EMC effect should continue to increase for all measured nuclei. This saturation / no-saturation is a non-trivial prediction of our model that is supported by the data.

In the per-neutron correlation, the proton-rich $^3$He point is far below the simple straight line, while the neutron-rich Fe and Pb points are above it. In the per-proton correlation, the proton-rich $^3$He point is below the simple straight line for $N = Z$ nuclei, while the increasingly neutron-rich heavy nuclei are above it. These features of the data are all well-described by our SRC-driven EMC model.

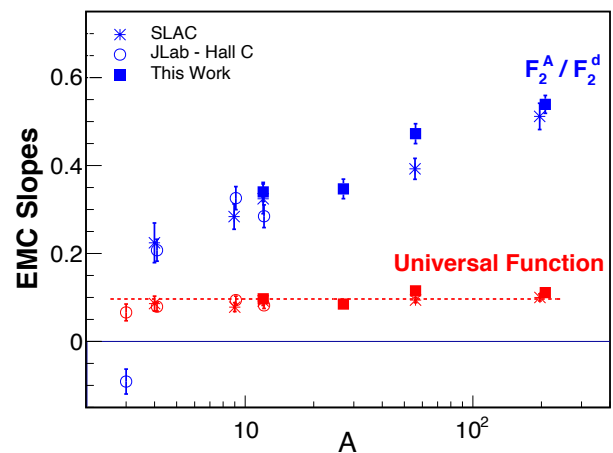

**Fig 3 | EMC and universal modification function slopes.** The slopes of the EMC effect for different nuclei from Fig. 2a (blue) and of the universal function from Fig. 2b (red). The error bars shown include the fit uncertainties at the 1σ or 68% confidence level.

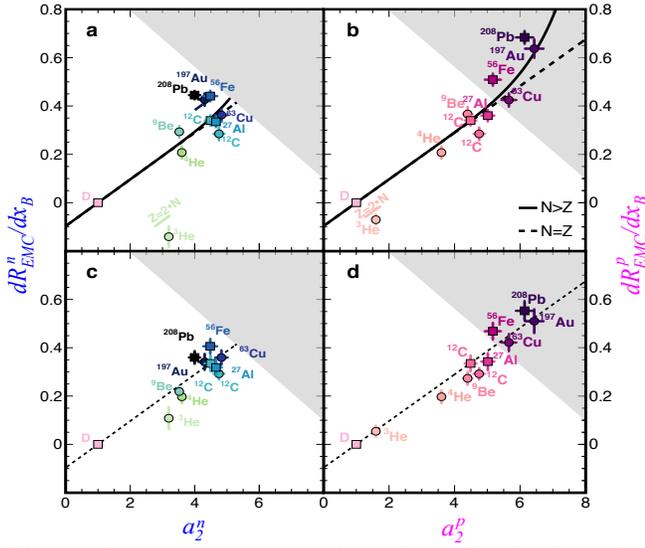

**Fig 4 | Growth and saturation of the EMC effect for protons and neutrons.** The (a) per-neutron and (b) per-proton strength of the EMC effect versus the corresponding per-neutron and per-proton number of SRC pairs. New data are shown by squares and existing data by circles. The dashed line shows the results of Eq. 2 using the universal modification function shown in Fig. 2 for symmetric N = Z nuclei. The solid line shows the same results for the actual nuclei. The gray region shows the effects of per-neutron saturation. (c) and (d): the same, but with isoscalar corrections. The error bars on the symbols show both statistical and systematic uncertainties, both at the 1σ or 68% confidence level.

To conclude, the association of the EMC effect with SRC pairs implies that it is a dynamical effect. Most of the time, nucleons bound in nuclei have the same internal structure as that of free nucleons. However, for short time intervals when two nucleons form a temporary high local-density SRC pair, their internal structure is briefly modified. When the two nucleons disassociate, their internal structure again becomes similar to that of free nucleons. This dynamical picture differs significantly from the traditional static modification in the nuclear mean-field, previously proposed as an explanation for the EMC effect.

The new universal modification function presented here has implications for our understanding of fundamental aspects of Quantum Chromodynamics (QCD). For example, the study of the ratio of the d-quark to u-quark population in a free nucleon as $x_B \to 1$ offers a stringent test of symmetry-breaking mechanisms in QCD. This can be extracted from measuring the free proton to neutron structure-function ratio. However, the lack of a free neutron target forces the use of proton and deuterium DIS data, which requires corrections for the deuteron EMC effect to extract the free neutron. The universal SRC modification function presented here does just that, in a data-driven manner, see Extended Data Fig. 1.

Turning to neutron-rich nuclei, the larger proton EMC effect has several implications. As the proton has two u-quarks and one d-quark while the neutron has two d-quarks and one u-quark, the larger average modification of the protons' structure implies a larger average modification of the distribution of u-quarks in the nucleus as compared to d-quarks. This will affect DIS charge-changing neutrino interactions, because neutrinos (ν) scatter preferentially from d-quarks and anti-neutrinos ($\bar{\nu}$) from u-quarks. Different modifications to d and u quark distributions will cause a difference in the ν and $\bar{\nu}$ cross-sections in asymmetric nuclei, which could then be misinterpreted as a sign of physics beyond the standard model or of CP-violation. One example of this is the NuTeV experiment, which extracted an anomalous value of the standard-model Weinberg mixing angle from ν and $\bar{\nu}$-nucleus DIS on iron. Ref. [30] pointed out that this anomaly could be due to differences between the proton and the neutron caused by mean-field effects. Our model provides an alternative mechanism. Similarly, the future DUNE experiment will use high-energy ν and $\bar{\nu}$ beams incident on the asymmetric nucleus $^{40}$Ar to look for differences in ν and $\bar{\nu}$ oscillations as a possible mechanism for explaining the matter-antimatter asymmetry. They will therefore also need to take the larger proton EMC effect into account to avoid similar anomalies.

**Acknowledgements** We acknowledge the efforts of the staff of the Accelerator and Physics Divisions at Jefferson Lab that made this experiment possible. The analysis presented here was carried out as part of the Jefferson Lab Hall B Data-Mining project supported by the U.S. Department of Energy (DOE). The research was supported also by the National Science Foundation, the Israel Science Foundation, the Chilean Comisión Nacional de Investigación Científica y Tecnológica, the French Centre National de la Recherche Scientifique and Commissariat a l'Energie Atomique, the French-American Cultural Exchange, the Italian Istituto Nazionale di Fisica Nucleare, the National Research Foundation of Korea, and the UKs Science and Technology Facilities Council. The research of M.S. was supported by the U.S. Department of Energy, Office of Science, Office of Nuclear Physics, under Award No. DE-FG02- 93ER40771. Jefferson Science Associates operates the Thomas Jefferson National Accelerator Facility for the DOE, Office of Science, Office of Nuclear Physics under contract DE-AC05-06OR23177.

**Author Contributions** The CEBAF Large Acceptance Spectrometer was designed and constructed by the CLAS Collaboration and Jefferson Lab. Data acquisition, processing and calibration, Monte Carlo simulations of the detector and data analyses were performed by a large number of CLAS Collaboration members, who also discussed and approved the scientific results. The analysis presented here was performed by B.S. and A.S. with input from S.G., O.H., E.P., and L.B.W., and reviewed by the CLAS collaboration.

**Author Information** Reprints and permissions information is available at www.nature.com/reprints. The authors declare no competing financial interests. Readers are welcome to comment on the online version of the paper. Publisher's note: Springer Nature remains neutral

**The CLAS Collaboration:** B. Schmookler,[1] M. Duer,[2] A. Schmidt,[1] O. Hen,[1] S. Gilad,[1] E. Piasetzky,[2] M. Strikman,[3] L.B. Weinstein,[4] S. Adhikari,[5] M. Amaryan,[4] A. Ashkenazi,[1] H. Avakian,[6] J. Ball,[7] I. Balossino,[8] L. Barion,[8] M. Battaglieri,[9] A. Beck,[1] I. Bedlinskiy,[10] A.S. Biselli,[11] S. Boiarinov,[6] W.J. Briscoe,[12] W.K. Brooks,[6,13] V.D. Burkert,[6] D.S. Carman,[6] A. Celentano,[9] G. Charles,[4] T. Chetry,[14] G. Ciullo,[8,15] E. Cohen,[2] P.L. Cole,[6,16,17] V. Crede,[18] R. Cruz-Torres,[1] A. D'Angelo,[19,38] N. Dashyan,[21] E. De Sanctis,[22] R. De Vita,[9] A. Deur,[6] C. Djalali,[47] R. Dupre,[23] H. Egiyan,[6] L. El Fassi,[24] L. Elouadrhiri,[6] P. Eugenio,[18] G. Fedotov,[14] R. Fersch,[25,26] A. Filippi,[20] G. Gavalian,[6] G.P. Gilfoyle,[27] F.X. Girod,[6] E. Golovatch,[28] R.W. Gothe,[47] K.A. Griffioen,[26] M. Guidal,[23] L. Guo,[5,6] H. Hakobyan,[13,21] C. Hanretty,[6] N. Harrison,[6] F. Hauenstein,[4] K. Hicks,[14] D. Higinbotham,[6] M. Holtrop,[29] C.E. Hyde,[4] Y. Ilieva,[12,47] D.G. Ireland,[30] B.S. Ishkhanov,[28] E.L. Isupov,[28] H-S. Jo,[31] S. Johnston,[32] S. Joosten,[33] M.L. Kabir,[24] D. Keller,[34] G. Khachatryan,[21] M. Khachatryan,[4] M. Khandaker,[39] A. Kim,[35] W. Kim,[31] A. Klein,[4] F.J. Klein,[17] I. Korover,[44] V. Kubarovsky,[6] S.E. Kuhn,[4] S.V. Kuleshov,[10,13] L. Lanza,[19] G. Laskaris,[1] P. Lenisa,[8] K. Livingston,[30] I.J.D. MacGregor,[30] N. Markov,[35] B. McKinnon,[30] S. Mey-Tal Beck,[1] T. Mineeva,[13] M. Mirazita,[22] V. Mokeev,[6,28] R.A. Montgomery,[30] C. Munoz Camacho,[23] B. Mustpha,[5] S. Niccolai,[23] M. Osipenko,[9] A.I. Ostrovidov,[18] M. Paolone,[33] R. Paremuzyan,[29] K. Park,[6,31] E. Pasyuk,[6,36] M. Patsyuk,[1] O. Pogorelko,[10] J.W. Price,[37] Y. Prok,[4,34] D. Protopopescu,[30] M. Ripani,[9] D. Riser,[35] A. Rizzo,[19,38] G. Rosner,[30] P. Rossi,[6,22] F. Sabatié,[7] C. Salgado,[39] R.A. Schumacher,[40] E.P. Segarra,[1] Y.G. Sharabian,[6] I.U. Skorodumina,[28,47] D. Sokhan,[30] N. Sparveris,[33] S. Stepanyan,[6] S. Strauch,[12,47] M. Taiuti,[9,41] J.A. Tan,[31] M. Ungaro,[6,42] H. Voskanyan,[21] E. Voutier,[23] D. Watts,[43] X. Wei,[6] M. Wood,[45] N. Zachariou,[43] J. Zhang,[34] Z.W. Zhao,[4,46] and X. Zheng[34]

[1]Massachusetts Institute of Technology, Cambridge, MA 02139
[2]Tel Aviv University, Tel Aviv, Israel
[3]Pennsylvania State University, University Park, PA, 16802
[4]Old Dominion University, Norfolk, Virginia 23529
[5]Florida International University, Miami, Florida 33199
[6]Thomas Jefferson National Accelerator Facility, Newport News, Virginia 23606
[7]IRFU, CEA, Université Paris-Saclay, F-91191 Gif-sur-Yvette, France
[8]INFN, Sezione di Ferrara, 44100 Ferrara, Italy
[9]INFN, Sezione di Genova, 16146 Genova, Italy
[10]Institute of Theoretical and Experimental Physics, Moscow, 117259, Russia
[11]Fairfield University, Fairfield, Connecticut 06824, USA
[12]The George Washington University, Washington, DC 20052
[13]Universidad Técnica Federico Santa María, Casilla 110-V Valparaíso, Chile
[14]Ohio University, Athens, Ohio 45701
[15]Universita' di Ferrara, 44121 Ferrara, Italy
[16]Idaho State University, Pocatello, Idaho 83209
[17]Catholic University of America, Washington, D.C. 20064
[18]Florida State University, Tallahassee, Florida 32306
[19]INFN, Sezione di Roma Tor Vergata, 00133 Rome, Italy
[20]INFN, Sezione di Torino, 10125 Torino, Italy
[21]Yerevan Physics Institute, 375036 Yerevan, Armenia
[22]INFN, Laboratori Nazionali di Frascati, 00044 Frascati, Italy
[23]Institut de Physique Nucléaire, CNRS/IN2P3 and Université Paris Sud, Orsay, France
[24]Mississippi State University, Mississippi State, MS 39762-5167
[25]Christopher Newport University, Newport News, Virginia 23606
[26]College of William and Mary, Williamsburg, Virginia 23187-8795
[27]University of Richmond, Richmond, Virginia 23173
[28]Skobeltsyn Institute of Nuclear Physics, Lomonosov Moscow State University, 119234 Moscow, Russia
[29]University of New Hampshire, Durham, New Hampshire 03824-3568
[30]University of Glasgow, Glasgow G12 8QQ, United Kingdom
[31]Kyungpook National University, Daegu 41566, Republic of Korea
[32]Argonne National Laboratory, Argonne, Illinois 60439
[33]Temple University, Philadelphia, PA 19122
[34]University of Virginia, Charlottesville, Virginia 22901
[35]University of Connecticut, Storrs, Connecticut 06269
[36]Arizona State University, Tempe, Arizona 85287-1504
[37]California State University, Dominguez Hills, Carson, CA 90747
[38]Universita' di Roma Tor Vergata, 00133 Rome Italy
[39]Norfolk State University, Norfolk, Virginia 23504
[40]Carnegie Mellon University, Pittsburgh, Pennsylvania 15213
[41]Universit`a di Genova, Dipartimento di Fisica, 16146 Genova, Italy.
[42]Rensselaer Polytechnic Institute, Troy, New York 12180-3590
[43]University of York, Heslington, York YO10 5DD, United Kingdom
[44]Nuclear Research Centre Negev, Beer-Sheva, Israel
[45]Canisius College, Buffalo, NY 14208, USA
[46]Duke University, Durham, North Carolina 27708-0305
[47]University of South Carolina, Columbia, South Carolina 29208

# Methods

**Experimental setup and electron identification.** CLAS used a toroidal magnetic field with six sectors of drift chambers, scintillation counters, Cerenkov counters and electromagnetic calorimeters to identify electrons and reconstruct their trajectories [23].

The experiment used a specially designed double target setup, consisting of a 2-cm long cryo-target cell, containing liquid deuterium, and a solid target [24]. The cryo-target cell and solid target were separated by 4 cm, with a thin isolation foil between them. Both targets and the isolation foil were kept in the beam line simultaneously. This allowed for an accurate measurement of cross-section ratios for nuclei relative to deuterium. A dedicated control system was used to position one of six different solid targets (thin and thick Al, Sn, C, Fe, and Pb, all in natural abundance) at a time during the experiment. The main data collected during the experiment was for a target configuration of deuterium + C, Fe, or Pb and also for an empty cryo-target cell with the thick Al target.

We identified electrons by requiring that the track originated in the liquid deuterium or solid targets, produced a large enough signal in the Cerenkov counter, and deposited enough energy in the Electromagnetic Calorimeter, see [21, 22] for details.

**Vertex reconstruction.** Electrons scattering from the solid and cryo-targets were selected using vertex cuts with a resolution of several mm (depending on the scattering angle), which is sufficient to separate the targets which are 4 cm apart [21]. We considered events with reconstructed electron vertex up to 0.5 cm outside the 2 cm long cryo-target to originate from the deuterium. Similarly, for the solid target, we considered events with reconstructed electron vertex up to 1.5 cm around it.

**Background subtraction.** There are two main sources of background in the measurement: (1) electrons scattering from the Al walls of the cryo-target cell, (2) electrons scattering from the isolation foil between the cryo-target and solid target. When the vertex of these electrons is reconstructed within the region of the deuterium target, they falsely contribute to the cross section associated with the deuterium target. Data from measurements done using an empty cryo-target is used to subtract these contributions. In the case of QE scattering, at $x_B > 1$, these measurements do not have enough statistics to allow for a reliable background subtraction. We therefore require QE deuterium electrons to be reconstructed in the inner 1-cm of the 2-cm long cryo-target. This increases the reliability of the background subtraction but reduces the deuterium statistics by a factor of two.

Data from runs with a full cryo-target and no solid target were used to subtract background from electron scattering events with a reconstructed vertex in the solid-target region, originating from the isolation foil or the cryo-target.

To increase statistics, the analysis combined all deuterium data, regardless of the solid target placed with it in the beam line. We only consider runs where the electron scattering rate from the cryo-target deviated by less than 4% from the average.

The systematic uncertainties associated with the vertex cuts, target wall subtraction, and combination of deuterium data from different runs are described in the Supplemental Materials, section 2.

**Data Availability:** The raw data from this experiment are archived in Jefferson Lab's mass storage silo.

# Extended Data

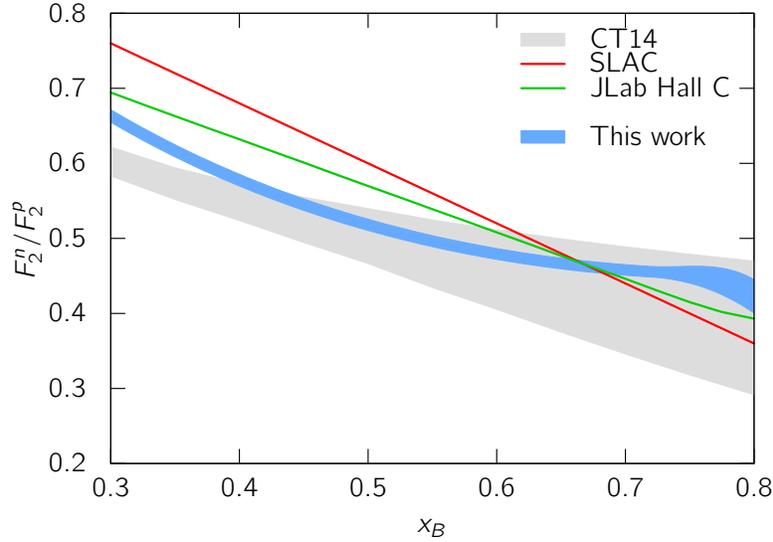

**Extended Data Fig 1 | $F_2^n/F_2^p$ Models.** The ratio of neutron to proton structure functions, $F_2^n/F_2^p$, derived from the SRC-driven EMC model (blue band), assumed in the isoscalar corrections of Refs. [9] (red line) and [10] (green line), and derived in the CT14 global fit, shown here for $Q^2 = 10$ GeV$^2$ (gray band). The large spread among the various models shows the uncertainty in $F_2^n$, a key ingredient in the isoscalar corrections previously applied to the EMC effect data

**Extended Data Table I: | SRC Scaling Coefficients.** Per-nucleon ($a_2$), per-proton ($a_2^p$), and per-neutron ($a_2^n$) SRC scale factors for nucleus $A$ relative to deuterium. The 1σ or 68% confidence level uncertainties shown include the fit uncertainties.

| Nucleus | This work | | | Ref. [5] | | |
|---|---|---|---|---|---|---|
| | $a_2$ | $a_2^p$ | $a_2^n$ | $a_2$ | $a_2^p$ | $a_2^n$ |
| $^3$He | | | | 2.13±0.04 | 1.60±0.03 | 3.20±0.06 |
| $^4$He | | | | 3.60±0.10 | 3.60±0.10 | 3.60±0.10 |
| $^9$Be | | | | 3.91±0.12 | 4.40±0.14 | 3.52±0.11 |
| $^{12}$C | 4.49±0.17 | 4.49±0.17 | 4.49±0.17 | 4.75±0.16 | 4.75±0.16 | 4.75±0.16 |
| $^{27}$Al | 4.83±0.18 | 5.02±0.19 | 4.66±0.17 | | | |
| $^{56}$Fe | 4.80±0.22 | 5.17±0.24 | 4.48±0.21 | | | |
| $^{63}$Cu | | | | 5.21±0.20 | 5.66±0.22 | 4.83±0.19 |
| $^{197}$Au | | | | 5.16±0.22 | 6.43±0.27 | 4.31±0.18 |
| $^{208}$Pb | 4.84±0.20 | 6.14±0.25 | 3.99±0.17 | | | |

**Extended Data Table II: | EMC Slopes.** Slopes of non isoscalar-corrected $F_2^A/F_2^d$ ($dR_{EMC}/dx_B$) and the universal function, shown in Figs. 2a and 2b of the main paper, respectively. The SLAC data is from [9] and the JLab Hall C data is from [10]. The slopes are obtained from a linear fit of the data for $0.25 \leq x_B \leq 0.7$. The 1σ or 68% confidence level uncertainties shown include the fit uncertainties.

| Nucleus | $dR_{EMC}/dx_B$ | | | Universal Function Slope | | |
|---|---|---|---|---|---|---|
| | JLab Hall C | SLAC | This Work | JLab Hall C | SLAC | This Work |
| $^3$He | 0.091±0.028 | | | -0.066±0.019 | | |
| $^4$He | -0.207±0.025 | -0.222±0.045 | | -0.080±0.010 | -0.086±0.017 | |
| $^9$Be | -0.326±0.026 | -0.283±0.028 | | -0.094±0.009 | -0.078±0.010 | |
| $^{12}$C | -0.285±0.026 | -0.322±0.033 | -0.340±0.022 | -0.082±0.007 | -0.092±0.010 | -0.097±0.006 |
| $^{27}$Al | | | -0.347±0.022 | | | -0.086±0.006 |
| $^{56}$Fe | | -0.391±0.025 | -0.472±0.023 | | -0.094±0.006 | -0.115±0.006 |
| $^{63}$Cu | | -0.391±0.025 | | | -0.094±0.006 | |
| $^{197}$Au | | -0.511±0.030 | | | -0.100±0.008 | |
| $^{208}$Pb | | | -0.539±0.020 | | | -0.111±0.005 |

**Extended Data Table III: | Per nucleon, per-proton, and per-neutron EMC Slopes.** Per-nucleon ($dR_{EMC}/dx_B$) per-proton ($dR_{EMC}^p/dx_B$) and per-neutron ($dR_{EMC}^n/dx_B$) EMC slopes from the current and previous works, used in Fig. 4 of the main paper. The previous data shows the JLab Hall C results [10] for light nuclei ($A \leq 12$) and the SLAC results [9] for heavier nuclei. The 1σ or 68% confidence level uncertainties shown include the fit uncertainties.

| Nucleus | This work | | | Previous Data | | |
|---|---|---|---|---|---|---|
| | $dR_{EMC}/dx_B$ | $dR_{EMC}^p/dx_B$ | $dR_{EMC}^n/dx_B$ | $dR_{EMC}/dx_B$ | $dR_{EMC}^p/dx_B$ | $dR_{EMC}^n/dx_B$ |
| $^3$He | | | | 0.091±0.028 | 0.068±0.021 | 0.137±0.041 |
| $^4$He | | | | -0.207±0.025 | -0.207±0.025 | -0.207±0.025 |
| $^9$Be | | | | -0.326±0.026 | -0.367±0.029 | -0.293±0.024 |
| $^{12}$C | -0.340±0.022 | -0.340±0.022 | -0.340±0.022 | -0.285±0.026 | -0.285±0.026 | -0.285±0.026 |
| $^{27}$Al | -0.347±0.022 | -0.360±0.023 | -0.335±0.021 | | | |
| $^{56}$Fe | -0.472±0.023 | -0.509±0.024 | -0.441±0.021 | -0.391±0.025 | -0.421±0.027 | -0.365±0.023 |
| $^{63}$Cu | | | | -0.391±0.025 | -0.425±0.027 | -0.362±0.023 |
| $^{197}$Au | | | | -0.511±0.030 | -0.637±0.037 | -0.427±0.025 |
| $^{208}$Pb | -0.539±0.020 | -0.684±0.026 | -0.445±0.017 | | | |

# Supplementary Materials for: Modified Structure of Protons and Neutrons in Correlated Pairs

## I. CROSS-SECTION RATIO EXTRACTION

Inclusive $(e, e')$ cross sections are differential in two variables. We follow the typical convention by choosing $x_B$ and $Q^2$. We extract ratios of cross sections for nuclei relative to deuterium as a function of $x_B$, integrated over $Q^2$. As CLAS has a large acceptance (as seen in Fig. 1), the integration over $Q^2$ covers a wide range of about $1.5 - 5$ GeV$^2$. However, as the EMC and QE ratios are $Q^2$ independent this is not a limitation [1–5, 12].

The cross-section extraction is done by weighting each measured event to correct for experimental effects as follows

$$weight = \frac{RC \times CC}{NORM \times ACC} \times BC \times ISO, \qquad (1)$$

where NORM is the experimental luminosity (beam charge times target thickness times the experimental live time), ACC is the acceptance correction and bin-migration factor, RC is the radiative correction factor, CC is the Coulomb correction factor, BC is the bin-centering correction and ISO is the isoscalar correction which can be applied to the $x_B < 1$ (DIS) data. (We include ISO in Eq. 1 for completeness, since isoscalar corrections were applied to previously published data, but we chose to omit this term for the data presented here.) These corrections and their associated systematic uncertainties are discussed in detail below. The resulting cross-section ratios and their uncertainties are listed in Tables I and II.

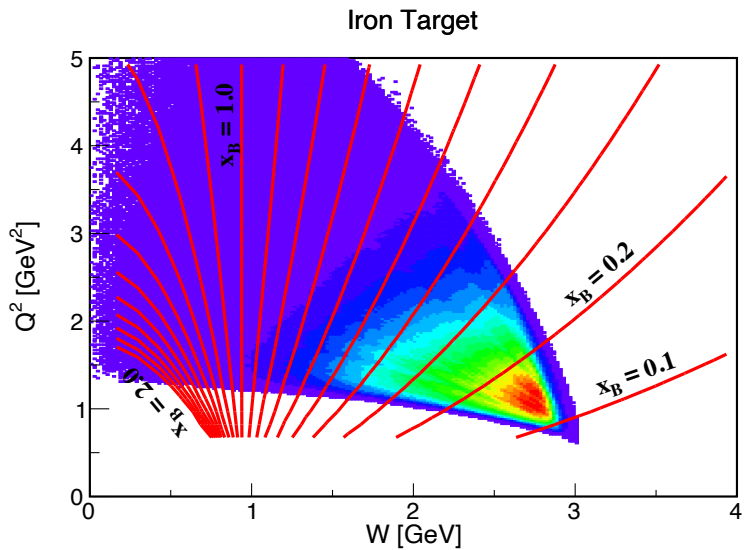

**Fig. 1**: | **CLAS $(e, e')$ Phase Space.** CLAS $(e, e')$ phase space in terms of $Q^2$ vs. $W$. The color scale indicates the measured event yield. The solid lines mark $Q^2$-$W$ combinations leading to fixed values of $x_B$.

**Model cross section:** The application of the correction factors used in Eq. 1 requires a model for both the Born and radiative cross section in our kinematical phase space of interest. We use here the code INCLUSIVE [6] that was used also in previous analyses [4, 5] and well reproduces the measured data of this work (see Figs. 2 and 3). The model cross sections are generated on a fine two-dimensional grid of $x_B$ and $Q^2$ and are linearly interpolated to determine the model cross section at any location between the grid points.

**Acceptance Corrections (ACC):** As the liquid deuterium and solid targets were placed at slightly different locations along the beam line, the detector acceptance for scattered electrons from each target is slightly different. This difference affects the measured relative yield and thus needs to be corrected for. In addition, the detector momentum and scattering angle reconstruction resolution introduces bin migration. The latter occurs when a particle with a certain momentum and angle is reconstructed with a slightly different momentum and angle and therefore is assigned to an incorrect $x_B$ and $Q^2$ bin.

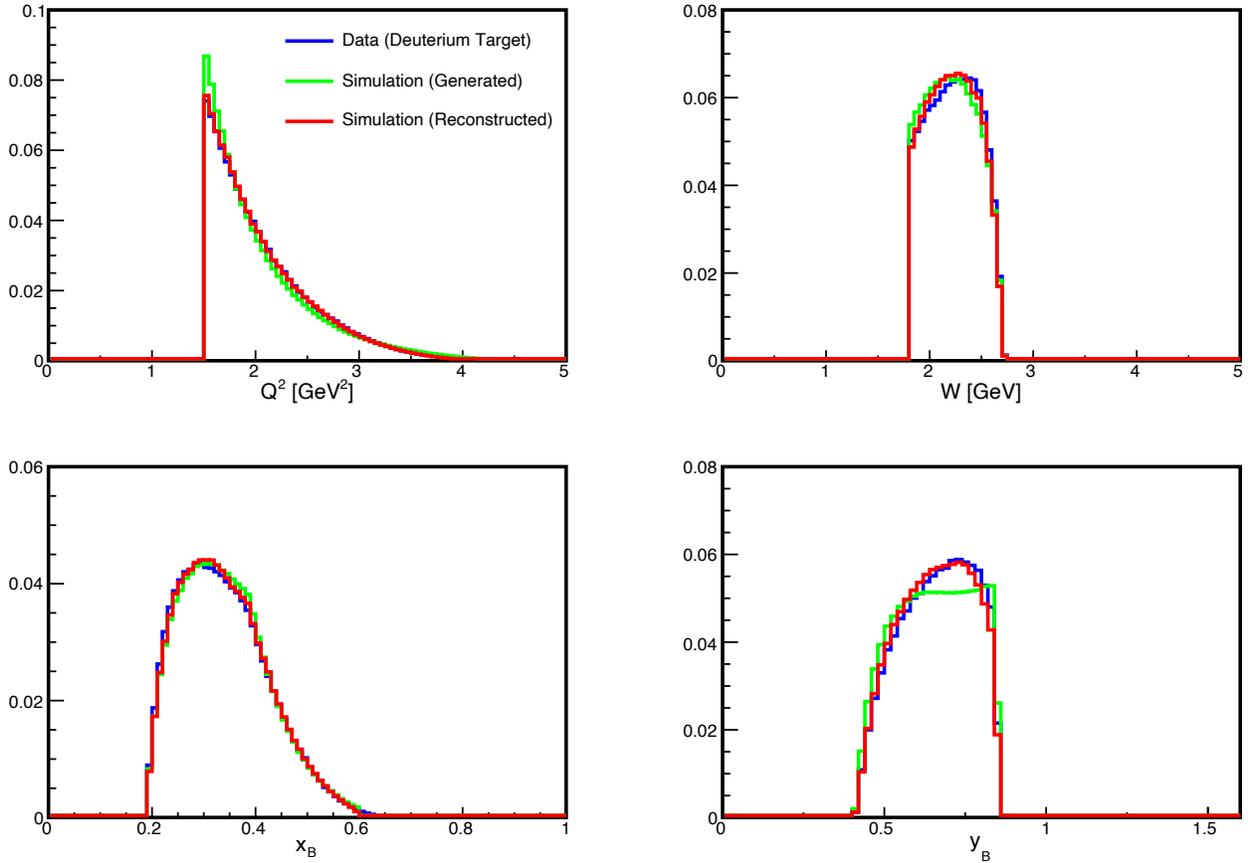

**Fig. 2**: | **Agreement Between Model Cross section and DIS Data.** Comparison of the shape of the measured DIS event yield (blue) with the simulated yields before (green) and after (red) passing through the CLAS detector acceptance simulation. All distributions are normalized to the same integral. Events shown are for DIS kinematics, after application of the $W \geq 1.8$ GeV, $Q^2 \geq 1.5$ GeV$^2$, and $Y \leq 0.85$ event selection cuts.

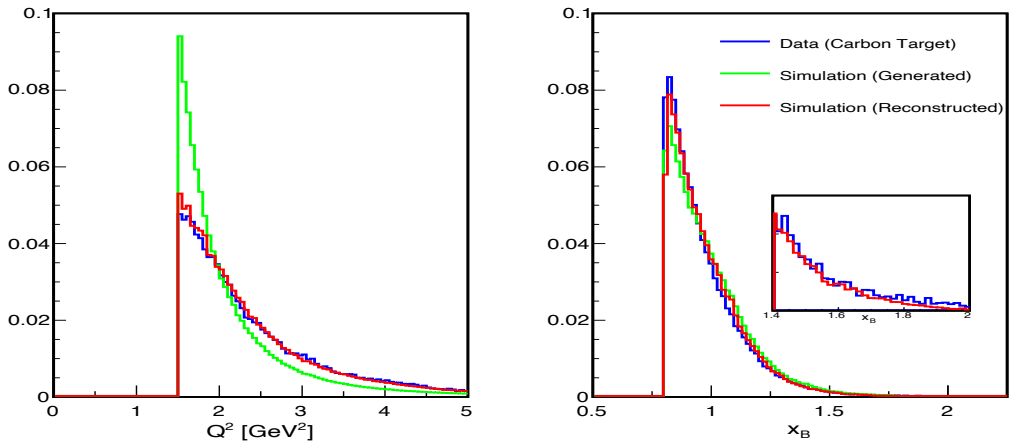

**Fig. 3**: | **Agreement Between Model Cross section and QE Data.** Same as Fig. 2, but for the selected QE events.

We determined the combined acceptance and bin-migration corrections using the CLAS Monte Carlo simulation as follows: we generated electrons uniformly in solid angle and energy, with vertices either in the solid target or along the liquid target. We then passed these events through the standard CLAS simulation chain, and weighted each event by its radiative model cross section, $\sigma_{Rad}(x_{gen}, Q^2_{gen})$ where $(x_{gen}, Q^2_{gen})$ are the kinematics of the generated electron. For the QE data, we finely binned the simulated events in $Q^2$ and $x_B$. For the DIS data, $Q^2$ and $W$ bins were used because kinematic cuts are applied to these variables. For each bin, the combined acceptance and bin-migration correction factor is defined as

$$ACC = \frac{\Sigma_{reconstructed} \sigma_{rad}(x_{gen}, Q^2_{gen})}{\Sigma_{generated} \sigma_{rad}(x_{gen}, Q^2_{gen})}, \tag{2}$$

where $\Sigma_{generated}$ refers to the sum over all generated electrons in that bin, and $\Sigma_{reconstructed}$ refers to the sum over all generated electrons that were detected and reconstructed by CLAS in that bin. The numerator includes events that migrated in (i.e., were generated with $(x_B, Q^2)$ outside the bin, but were reconstructed with $(x_B, Q^2)$ inside the bin) and excludes events that migrated out (i.e., were generated with $(x_B, Q^2)$ in that bin but were reconstructed with $(x_B, Q^2)$ outside the bin). This acceptance correction factor was then applied, event-by-event, to the measured data using the reconstructed electron kinematics to determine the appropriate bin.

**Radiative Corrections (RC):** Radiative corrections are applied to obtain the underlying Born cross section from the measured radiated data. This is done by using the cross-section model, calculated without and with radiative effects. The latter is done using the prescription of Ref. [7]. For each event, we calculated the radiative correction as

$$RC = \frac{\sigma_{Born}(x_B, Q^2)}{\sigma_{Rad}(x_B, Q^2)}, \tag{3}$$

where the Born and radiated cross sections are calculated at the kinematics of each event.

**Coulomb Corrections (CC):** As electrons scatter from a nucleus, they are first accelerated and then decelerated by the electric field of the nucleus. This means that the measured beam energy and scattered momentum are not equivalent to the values they have at the reaction vertex. Using the Effective Momentum Approximation (EMA) [8], both the initial and final electrons energies at the reaction vertex are higher by an amount $\Delta E$ as compared to their measured values. The calculation of $\Delta E$ for our beam energy and targets was done in Ref. [9].

The Coulomb Correction factors are given by the ratio of the cross section calculated at the Coulomb shifted and unshifted kinematics times a focusing factor as follows

$$CC = \frac{\sigma_{Born}(E, E', \theta)}{\sigma_{Born}(E + \Delta E, E' + \Delta E, \theta)} (E/(E + \Delta E))^2, \tag{4}$$

where $E, E'$, and $\theta$ are at the kinematics of each event.

**Isoscalar Corrections (ISO):** Previous studies of the EMC effect [2, 10, 11] included an isoscalar correction factor to account for the unequal number of protons and neutrons in many nuclei. This correction factor adjusts the measured per-nucleon cross section for nucleus $A$ to a new value which represents the per-nucleon cross section for a nucleus $A$ with equal numbers of neutrons and protons. This correction factor is given by

$$ISO = \frac{\frac{A}{2}(1 + \frac{\sigma_n}{\sigma_p})}{Z + N\frac{\sigma_n}{\sigma_p}}, \tag{5}$$

where $\sigma_n$ and $\sigma_p$ are the elementary electron-neutron and electron-proton cross sections, respectively. The lack of a free neutron target makes this correction strongly model-dependent (see Extended Data Fig. 1). Therefore, we have not applied isoscalar correction in this work for either DIS and QE cross-section ratios, except for the bottom panel of the paper Fig. 4 where we used $\frac{\sigma_n}{\sigma_p}$ extracted from our data and the universal modification function.

**Bin Centering Correction (BC):** As the cross sections fall rapidly as a function of $x_B$, binning the data could bias the extracted values of the cross-section ratio in a bin-width dependent manner. Bin-centering corrections are therefore used to move each event from its actual location in the $(x_B, Q^2)$ bin to the center of the bin as

$$BC = \frac{\sigma_{born}(x_{center}, Q^2_{event})}{\sigma_{born}(x_{event}, Q^2_{event})}, \tag{6}$$

where $x_{event}$ is the measured $x_B$ of the event and $x_{center}$ is the value of the center of the $x_B$-bin that the event is associated with.

The DIS and QE cross-section ratios were extracted using bin width of $\Delta x_B = 0.013$ for DIS and $\Delta x_B = 0.043$ for QE (except for the three highest QE points that used wide bins of $\Delta x_B = 0.086$). As a sensitivity study we examined additional binnings of $\Delta x = 0.010, 0.020, 0.040$ for DIS and $\Delta x = 0.086$ for QE. The extracted EMC slopes and SRC scaling coefficients were not sensitive to the bin-width choice.

## II. SYSTEMATIC UNCERTAINTIES

The corrections and weighting factors used in the cross-section ratio extraction procedure described above introduce systematic uncertainties to the resulting cross-section ratios. Here we list each source of systematic uncertainty, how it was evaluated, and its magnitude. We consider both overall normalization and point-to-point uncertainties. The latter are added in quadrature to the statistical uncertainties of the cross-section ratio in each $x_B$ bin while the former are common normalization uncertainties for all $x_B$ bins of a given cross-section ratio. Tables III and IV list the resulting point-to-point and normalization uncertainties for DIS and QE cross-section ratios respectively. We also consider systematic uncertainties arising from the analysis procedure that impact the resulting EMC slopes and QE cross-section scaling coefficients. These are detailed below.

**Beam Charge and Time-Dependent Instabilities:** Since we combine all the deuterium runs when calculating the cross-section ratios, our absolute normalization is sensitive to changes in the beam charge monitoring devices, fluctuations in the cryo-target, and changes to the CLAS detector over the run period. This is estimated by examining the systematic changes in the normalized yield for the deuterium target from different runs. We find the distribution of the deviation from the mean to be normally distributed with a sigma of ±0.65%. We conservatively place a systematic normalization uncertainty of 1% on the cross-section ratio.

**Target Thickness and Vertex Cuts:** The uncertainty in the cryo-target thickness has been estimated to be 1.0%. The thicknesses of the solid targets were measured to about 1-micron accuracy, which corresponds to a relative uncertainty of 0.1 – 0.7%.

The cryo-target vertex cuts for DIS kinematics were 3 cm wide. We varied this cut by 0.25 cm and examined the change in the windows-subtracted yield in each $x_B$ bin to find a maximal change in the yield of 1.0%. In QE kinematics, we applied a 1 cm wide cut in the center of the cryo-target. The uncertainty due to this cut stems from the vertex reconstruction. To test this, we measured the reconstructed window locations for the empty target runs and found a maximal deviation of 1% from the ideal 2-cm target length.

The final systematic uncertainty in the cross-section ratios due to the normalization combines the cryo-target thickness, solid-target thickness, and vertex cut uncertainties. This gives a normalization uncertainty of 1.42 – 1.58% in both the DIS and QE regions.

In addition, we examine the sensitivity of the extracted EMC slopes to using a 1 cm wide vertex cut instead of a 3 cm wide cut for the DIS kinematics. This change mainly affects the background levels and is included as a systematic uncertainty on the measured slope.

**Acceptance Corrections and Bin Migration:** The statistical uncertainty of the acceptance correction factors in the DIS and QE regions in each two-dimensional bin are 0.75% and 3.0%, respectively. After summing the data into one-dimensional bins in $x_B$, it is reduced to 0.25% and 0.75% respectively. Since the acceptance correction factors are applied to the deuterium and solid target separately, the effect on the cross-section ratios are 0.35% and 1.06% for the DIS and QE regions, respectively, which we apply as a point-to-point systematic uncertainty. In addition, we place a 0.5% normalization uncertainty on the acceptance due to imperfections in the detector simulation.

Bin migration is corrected for by weighting the acceptance map using the model cross sections. The systematic uncertainty on this correction can be estimated by examining how much bin migration affects the final ratios if no correction were applied. We studied this by performing the acceptance corrections using the uniform generator, without weighting the events with the cross-section model. The difference in the measured EMC slopes and $a_2$ values when using the two types of acceptance maps are included as a systematic uncertainty on the EMC slopes and $a_2$ values.

**Radiative, Coulomb, and Bin-Centering Corrections:** Point-to-point uncertainties due to the radiative corrections can arise due to detector resolution and bin migration. We studied this effect for both DIS and QE regions by comparing the generated and reconstructed weighted simulation after applying acceptance corrections to the reconstructed events. Then we considered the average radiative correction in each bin using both the generated (i.e., the true correction) and the acceptance-corrected reconstructed (i.e., the used correction) events. We take the ratio of the true correction to the used correction to determine the size of the resolution effect. We see that the effect cancels to < 0.01% in the final cross-section ratio. Point-to-point uncertainties that are not due to the resolution are expected to cancel in the ratio [2] and are therefore not applied. The normalization uncertainty on the cross-section ratios due to radiative corrections is estimated to be 0.5% [2, 11].

Coulomb corrections use an energy shift calculated from the Coulomb potential, which has a 10% uncertainty. We study the impact of this on the Coulomb correction factors by recalculating them using a $\Delta E$ in Eq. 4 that is changed by 10%. For the DIS region, this changes the Coulomb correction factor by a maximum of only 0.1%. For the QE region, the factor changes by a maximum of 0.2% for carbon, 0.4% for aluminum, 0.7% for iron, and 1.0% for lead. Although there is some $x_B$ dependence to the change in the correction factor, they are correlated. Therefore, we conservatively apply the maximum change for each target as a normalization uncertainty.

Bin-centering systematic uncertainties are estimated by examining the difference in the resulting EMC slopes and $a_2$ values when applying the bin-centering corrections prior to all the other corrections in Eq. 1. Following previous work, we also place a 0.5% point-to-point uncertainty on the bin-centering correction factor.

**Kinematic Corrections:** For the QE case, we estimate that the maximum amount that the electron momentum may be reconstructed incorrectly is 20 MeV/c, using deuteron breakup measurements. To check the effect of this potential mis-reconstruction on the cross-section ratios, we examined the variation in the measured cross-section ratio when shifting the scattered electron momentum by 20 MeV/c. We find that the ratio changes between 0.2-0.3%. We therefore place a point-to-point uncertainty of 0.3% on this. For the DIS case, we applied momentum and polar-angle corrections using exclusive hydrogen measurements and do not place any uncertainty on these corrections.

### III. SRC SCALING COEFFICIENT EXTRACTION

The relative abundances of SRC pairs in nuclei is extracted from the measured per-nucleon QE cross-section ratios presented above. For $Q^2 > 1.5$ GeV$^2$ and $1.5 < x_B < 2$, the cross-section ratio of any nucleus relative to deuterium ($\sigma_A/\sigma_d$) shows scaling, i.e., it is flat as a function of $x_B$, see Fig. 1 in the main text. The value of the per-nucleon cross-section ratio, referred to here as $a_2$ or the SRC scaling coefficient, is often interpreted as a measure of the relative abundance of high-momentum nucleons in the measured nucleus relative to deuterium [3–5, 12, 13].

While traditionally normalized to the number of nucleons $A$ (i.e., per-nucleon), the cross-section ratio can be normalized to the number of protons $Z$ (i.e., per-proton), or neutrons $N$ (i.e., per-neutron) in the measured nuclei. These different normalizations allow obtaining the relative fraction of high-momentum nucleons out of all nucleons in the nucleus, or just the protons or neutrons. We mark these ratios by $a_2$, $a_2^p$ and $a_2^n$ respectively:

$$\begin{aligned} a_2 &= \frac{2}{A} \cdot \frac{\sigma_A(Q^2, x_B)}{\sigma_d(Q^2, x_B)}|_{Q^2>1.5, 1.5 \leq x_B \leq 2}, \\ a_2^p &= \frac{1}{Z} \cdot \frac{\sigma_A(Q^2, x_B)}{\sigma_d(Q^2, x_B)}|_{Q^2>1.5, 1.5 \leq x_B \leq 2}, \\ a_2^n &= \frac{1}{N} \cdot \frac{\sigma_A(Q^2, x_B)}{\sigma_d(Q^2, x_B)}|_{Q^2>1.5, 1.5 \leq x_B \leq 2}. \end{aligned} \quad (7)$$

Extended Data Table I lists the values and uncertainties of $a_2$, $a_2^p$ and $a_2^n$, extracted from measurements presented in this work and the world data compilation of Ref. [14], Table 1, column 6, based on the measurements of Refs. [4, 5, 13].

Eq. 1 in the main text uses $n_{SRC}^A$, the number of nucleons that are part of $np$-SRC pairs. In the SRC-driven EMC model this is given by [12]:

$$\begin{aligned} n_{SRC}^A &= A \cdot a_2 \cdot \frac{n_{SRC}^d}{2} \\ &= (Za_2^p + Na_2^n) \cdot \frac{n_{SRC}^d}{2}. \end{aligned} \quad (8)$$

### IV. DIS CROSS SECTIONS AND STRUCTURE FUNCTIONS

The DIS cross section for scattering a high-energy electron or muon from a nuclear target of mass $A$ depends on two structure functions, $F_1^A(x_B, Q^2)$ and $F_2^A(x_B, Q^2)$. At large enough momentum transfer, $F_1^A$ and $F_2^A$ are independent of $Q^2$ and describe the structure of the target nucleus. The ratio of DIS cross sections for nucleus $A$ and deuterium equals the ratio of the $F_2$ structure functions when the ratios of the absorption cross sections for longitudinal and transverse virtual photons are the same in nucleus $A$ and in deuterium. While this is typically assumed to be true, there are few measurements of this ratio in nuclei. See [1, 15] for details.

The EMC structure-function ratio is independent of $Q^2$ at relatively low $Q^2$. This was shown in [2] down to $Q^2 = 2$ GeV$^2$ and in our cut sensitivity study down to $Q^2 = 1.5$ GeV$^2$.

### V. EMC SLOPE EXTRACTION

We characterize the strength of the EMC effect for each nucleus as the slope [11] of the ratio of the per-nucleon DIS electron scattering cross-section ratio for that nucleus relative to deuterium, $dR_{EMC}/dx_B$ in the region $0.25 \leq x_B \leq 0.7$. Here we also calculate separately the slope of the DIS ratio per proton, $dR_{EMC}^p/dx_B$, and per neutron, $dR^n_{EMC}/dx_B$, similarly to Eq. 7 above only for DIS cross-section ratios. The resulting values are listed in Extended Data Table III and include both the new measurements presented in this work as well as the world-data compilation of Ref. [14] based on the measurements of Refs. [2, 11]. Notice that, as in Refs. [11, 16], by focusing on the $0.25 \leq x_B \leq 0.7$ region, the uncertainties are not meant to take into account possible effects of the anti-shadowing region at $x_B \approx 0.15$ and the Fermi motion region at $x_B > 0.75$ extending into the region of interest.

## VI. ANALYSIS OF PREVIOUS EMC DATA

Previous EMC data (from [2, 11]) have been reanalyzed to remove their isoscalar corrections. This was done by dividing the EMC ratios for asymmetric nuclei by Eq. 5. Each data-set was corrected using the $\sigma_n/\sigma_p$ parametrization used in its analysis, given by $\sigma_n/\sigma_p = 1 - 0.8 \cdot x_B$ for Ref. [2] and tabulated values for Ref. [11] (see Extended Data Fig. 1). Following [17], we multiply the $^3$He/$^2$H ratio of [11] by 1.03 for consistency with other data. It has no impact on the extracted EMC slopes.

## VII. SRC MODEL OF EMC RATIOS

The model presented in Eq. 1 in the main text can be used to predict the ratio of the per-nucleon structure functions for nucleus $A$ relative to deuterium (i.e., the EMC effect) as:

$$\frac{F_2^A/A}{F_2^d/2} = (a_2 - 2\frac{N}{A})(n^d_{SRC}\frac{\Delta F_2^p + \Delta F_2^n}{F_2^d}) \\ + 2 \cdot \frac{Z-N}{Z+N} \cdot \frac{F_2^p}{F_2^d} + 2\frac{N}{A}. \quad (9)$$

The same model can be used to predict the ratio of the per-proton and per-neutron EMC ratios (see Fig. 4 in the main text):

$$\frac{F_2^A/N}{F_2^d/1} = (a_2^n - 1)(n^d_{SRC}\frac{\Delta F_2^p + \Delta F_2^n}{F_2^d}) \\ + (\frac{Z}{N} - 1) \cdot \frac{F_2^p}{F_2^d} + 1, \\ \frac{F_2^A/Z}{F_2^d/1} = (a_2^p - \frac{N}{Z})(n^d_{SRC}\frac{\Delta F_2^p + \Delta F_2^n}{F_2^d}) \\ + (\frac{Z}{N} - 1) \cdot \frac{F_2^p}{F_2^d} + \frac{N}{Z}. \quad (10)$$

The theory prediction shown in Fig. 4 of the main text was obtained by calculating Eq. 10 for each nucleus and fitting the resulting slope for the per-proton and per-neutron ratios for $0.25 < x_B < 0.7$.

When self-consistent isoscalar corrections are applied, the $N/Z$ terms almost vanish, see Fig. 4.

As mentioned in the text, nucleon motion effects are incorporated into $\Delta F_2^p$ and $\Delta F_2^n$. This is a valid approximation since nucleon motion effects are proportional to kinetic energy, which is dominated by nucleons belonging to SRC pairs [12, 17, 18].

## VIII. THE EFFECT OF THREE-NUCLEON CORRELATIONS (3NC) AND TWO-BODY CURRENTS:

For the kinematics of the data reported in this work (i.e., $x_B < 2$), 3N-SRCs constitute a small correction to 2N-SRCs. Current estimates discuss a probability on the order of the 2N-SRC probability squared, which means its about an order of magnitude smaller contribution as compared with 2N-SRC.

Two-body currents manifest themselves as a change in the cross section ratios with $Q^2$. In DIS, the measured EMC effect ratios are observed to be independent of $Q^2$ for $2 \leq Q^2 \leq 40$ GeV$^2$ [2]. Hence the leading twist dominates in the ratio, and the virtual photon can be treated as if it interacts predominantly with individual quarks and antiquarks, not with two-body currents. The antiquark contribution is known to be very small for $x_B > 0.3$ for nucleons. Interactions with a meson (i.e., two-body) current would contribute to both quark and antiquark and would be observed as an enhancement of the antiquark distribution in nuclei at $x_B \approx 0.1$. This was tested by dedicated Drell-Yan pair production experiments performed at FNAL that did not observe such an effect. Thus two-body currents will be very small.

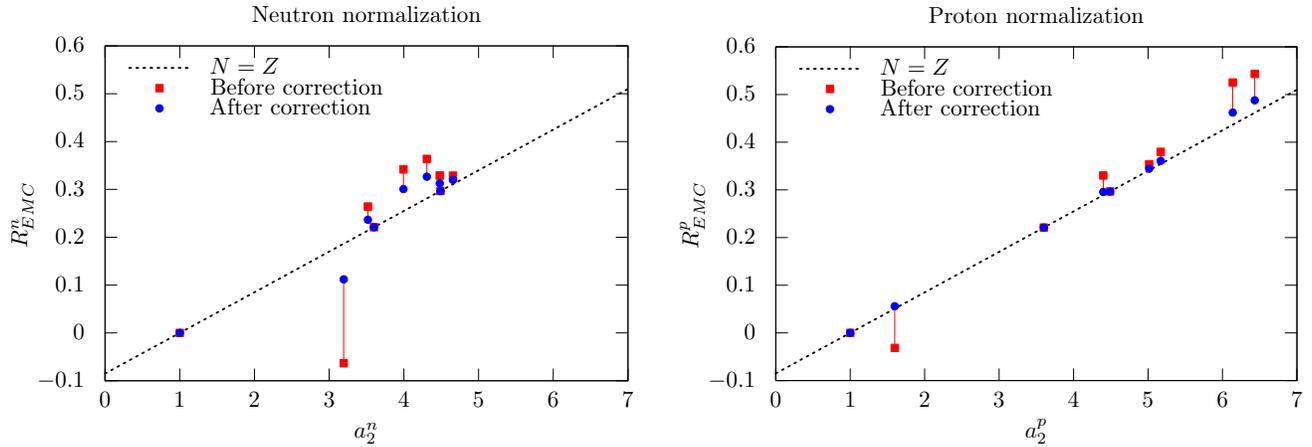

**Fig. 4**: **Effects of Isoscalar Corrections.** The per-neutron and per-proton EMC-slope predictions of Eq. 10 for the various nuclei shown in Fig. 4 of the main text, without (red squares) and with (blue circles) applying self-consistent isoscalar corrections.

In the QE region, two-body currents (Meson Exchange Currents and Isobar Configurations) are expected to be small at $x_B > 1.2$. This is confirmed experimentally by the fact that the cross-section ratios at $1.5 < x_B < 1.9$ do not depend on $Q^2$ as shown by this data at $Q^2 \approx 1.9$ GeV$^2$ and the previous JLab data at $Q^2 = 2.7$ GeV$^2$ [13].

**Table I:** | **DIS Cross-Section Ratios.** Tabulated values and uncertainties for the per-nucleon, non isoscalar-corrected $(e,e')$ DIS cross-section ratios for nuclei relative to deuterium as a function of $x_B$.

| $x_B$ | $\dfrac{\sigma_{\text{C}}/12}{\sigma_d/2}$ Norm: 1.81% | $\dfrac{\sigma_{\text{Al}}/27}{\sigma_d/2}$ Norm: 1.82% | $\dfrac{\sigma_{\text{Fe}}/56}{\sigma_d/2}$ Norm: 1.83% | $\dfrac{\sigma_{\text{Pb}}/208}{\sigma_d/2}$ Norm: 1.94% |
|---|---|---|---|---|
| 0.220 | $1.054 \pm 0.053$ | $1.001 \pm 0.050$ | $1.017 \pm 0.051$ | $1.016 \pm 0.051$ |
| 0.247 | $1.032 \pm 0.008$ | $1.002 \pm 0.008$ | $1.010 \pm 0.008$ | $0.999 \pm 0.008$ |
| 0.260 | $1.022 \pm 0.008$ | $0.995 \pm 0.008$ | $1.005 \pm 0.008$ | $0.988 \pm 0.008$ |
| 0.273 | $1.018 \pm 0.008$ | $0.998 \pm 0.008$ | $1.003 \pm 0.008$ | $0.982 \pm 0.008$ |
| 0.287 | $1.009 \pm 0.008$ | $0.996 \pm 0.008$ | $0.995 \pm 0.008$ | $0.975 \pm 0.008$ |
| 0.300 | $1.005 \pm 0.008$ | $0.993 \pm 0.008$ | $0.990 \pm 0.008$ | $0.967 \pm 0.008$ |
| 0.313 | $1.008 \pm 0.008$ | $0.989 \pm 0.008$ | $0.991 \pm 0.008$ | $0.964 \pm 0.008$ |
| 0.327 | $1.009 \pm 0.008$ | $0.994 \pm 0.008$ | $0.990 \pm 0.008$ | $0.964 \pm 0.008$ |
| 0.340 | $1.005 \pm 0.008$ | $0.990 \pm 0.008$ | $0.983 \pm 0.008$ | $0.958 \pm 0.008$ |
| 0.353 | $0.994 \pm 0.008$ | $0.973 \pm 0.008$ | $0.968 \pm 0.008$ | $0.945 \pm 0.008$ |
| 0.367 | $0.989 \pm 0.008$ | $0.970 \pm 0.008$ | $0.963 \pm 0.008$ | $0.937 \pm 0.008$ |
| 0.380 | $0.985 \pm 0.008$ | $0.967 \pm 0.008$ | $0.959 \pm 0.008$ | $0.931 \pm 0.007$ |
| 0.393 | $0.976 \pm 0.008$ | $0.959 \pm 0.008$ | $0.948 \pm 0.008$ | $0.919 \pm 0.007$ |
| 0.407 | $0.991 \pm 0.008$ | $0.974 \pm 0.008$ | $0.958 \pm 0.008$ | $0.931 \pm 0.008$ |
| 0.420 | $0.980 \pm 0.008$ | $0.964 \pm 0.008$ | $0.949 \pm 0.008$ | $0.914 \pm 0.007$ |
| 0.433 | $0.959 \pm 0.008$ | $0.942 \pm 0.008$ | $0.928 \pm 0.007$ | $0.896 \pm 0.007$ |
| 0.447 | $0.957 \pm 0.008$ | $0.943 \pm 0.008$ | $0.924 \pm 0.007$ | $0.896 \pm 0.007$ |
| 0.460 | $0.950 \pm 0.008$ | $0.932 \pm 0.008$ | $0.914 \pm 0.007$ | $0.880 \pm 0.007$ |
| 0.473 | $0.956 \pm 0.008$ | $0.940 \pm 0.008$ | $0.918 \pm 0.007$ | $0.886 \pm 0.007$ |
| 0.487 | $0.940 \pm 0.008$ | $0.920 \pm 0.008$ | $0.901 \pm 0.007$ | $0.872 \pm 0.007$ |
| 0.500 | $0.939 \pm 0.008$ | $0.925 \pm 0.008$ | $0.892 \pm 0.007$ | $0.861 \pm 0.007$ |
| 0.513 | $0.948 \pm 0.008$ | $0.924 \pm 0.009$ | $0.901 \pm 0.007$ | $0.861 \pm 0.008$ |
| 0.527 | $0.936 \pm 0.008$ | $0.901 \pm 0.009$ | $0.880 \pm 0.007$ | $0.843 \pm 0.008$ |
| 0.540 | $0.931 \pm 0.008$ | $0.905 \pm 0.009$ | $0.874 \pm 0.007$ | $0.839 \pm 0.008$ |
| 0.553 | $0.906 \pm 0.019$ | $0.873 \pm 0.019$ | $0.856 \pm 0.017$ | $0.812 \pm 0.017$ |
| 0.580 | $0.926 \pm 0.047$ | $0.919 \pm 0.046$ | $0.888 \pm 0.045$ | $0.812 \pm 0.041$ |

**Table II:** | **QE Cross-Section Ratios.** Tabulated values and uncertainties for the per-nucleon (e,e') QE cross-section ratios for nuclei relative to deuterium as a function of $x_B$.

| $x_B$ | $\dfrac{\sigma_{\rm C}/12}{\sigma_d/2}$<br>Norm: 1.82% | $\dfrac{\sigma_{\rm Al}/27}{\sigma_d/2}$<br>Norm: 1.85% | $\dfrac{\sigma_{\rm Fe}/56}{\sigma_d/2}$<br>Norm: 1.95% | $\dfrac{\sigma_{\rm Pb}/208}{\sigma_d/2}$<br>Norm: 2.18% |
|---|---|---|---|---|
| 0.821 | 1.335 ± 0.018 | 1.304 ± 0.018 | 1.278 ± 0.017 | 1.221 ± 0.017 |
| 0.864 | 1.140 ± 0.016 | 1.114 ± 0.016 | 1.087 ± 0.015 | 1.018 ± 0.014 |
| 0.907 | 0.777 ± 0.011 | 0.747 ± 0.011 | 0.727 ± 0.010 | 0.677 ± 0.010 |
| 0.950 | 0.557 ± 0.008 | 0.531 ± 0.008 | 0.517 ± 0.007 | 0.484 ± 0.007 |
| 0.992 | 0.509 ± 0.007 | 0.487 ± 0.007 | 0.474 ± 0.007 | 0.436 ± 0.006 |
| 1.036 | 0.660 ± 0.009 | 0.635 ± 0.010 | 0.610 ± 0.009 | 0.561 ± 0.008 |
| 1.079 | 0.928 ± 0.014 | 0.937 ± 0.015 | 0.885 ± 0.013 | 0.825 ± 0.013 |
| 1.121 | 1.278 ± 0.019 | 1.267 ± 0.021 | 1.224 ± 0.018 | 1.145 ± 0.018 |
| 1.164 | 1.686 ± 0.027 | 1.739 ± 0.031 | 1.704 ± 0.026 | 1.576 ± 0.026 |
| 1.207 | 2.152 ± 0.037 | 2.245 ± 0.044 | 2.145 ± 0.035 | 2.013 ± 0.037 |
| 1.250 | 2.651 ± 0.050 | 2.746 ± 0.059 | 2.613 ± 0.047 | 2.495 ± 0.050 |
| 1.293 | 3.128 ± 0.066 | 3.195 ± 0.079 | 3.067 ± 0.061 | 2.926 ± 0.066 |
| 1.336 | 3.604 ± 0.085 | 3.738 ± 0.103 | 3.552 ± 0.079 | 3.532 ± 0.089 |
| 1.379 | 4.002 ± 0.109 | 4.144 ± 0.133 | 3.992 ± 0.102 | 3.963 ± 0.115 |
| 1.421 | 4.362 ± 0.136 | 4.690 ± 0.171 | 4.544 ± 0.133 | 4.428 ± 0.147 |
| 1.464 | 4.634 ± 0.164 | 4.869 ± 0.203 | 4.920 ± 0.163 | 4.872 ± 0.184 |
| 1.507 | 4.209 ± 0.169 | 4.529 ± 0.212 | 4.490 ± 0.169 | 4.563 ± 0.194 |
| 1.550 | 4.501 ± 0.228 | 5.062 ± 0.288 | 4.684 ± 0.225 | 4.765 ± 0.252 |
| 1.593 | 4.289 ± 0.226 | 4.828 ± 0.291 | 4.590 ± 0.227 | 4.634 ± 0.256 |
| 1.636 | 4.368 ± 0.251 | 4.525 ± 0.307 | 4.701 ± 0.252 | 4.883 ± 0.294 |
| 1.679 | 4.610 ± 0.301 | 5.408 ± 0.406 | 5.088 ± 0.310 | 4.847 ± 0.337 |
| 1.721 | 4.644 ± 0.348 | 4.978 ± 0.431 | 5.188 ± 0.363 | 4.924 ± 0.389 |
| 1.786 | 4.951 ± 0.340 | 5.088 ± 0.398 | 5.245 ± 0.342 | 5.705 ± 0.405 |
| 1.871 | 5.107 ± 0.395 | 4.931 ± 0.453 | 5.553 ± 0.403 | 5.942 ± 0.481 |
| 1.957 | 5.527 ± 1.019 | 6.645 ± 1.303 | 5.477 ± 0.992 | 4.711 ± 0.893 |

**Table III:** | **DIS Systematic Uncertainties.** Systematic uncertainties in extraction of the DIS cross-section ratio.

| Source | Point-to-point (%) | Normalization (%) |
|---|---|---|
| Time-Dependent Instabilities | — | 1.0 |
| Target Thickness and Cuts | — | 1.42–1.58 |
| Acceptance Corrections | 0.6 (2,5) | — |
| Radiative Corrections | — | 0.5 |
| Coulomb Corrections | — | 0.1 |
| Bin-Centering Corrections | 0.5 | — |
| Total | 0.78 | 1.81–1.94 |

**Table IV:** | **QE Systematic Uncertainties.** Systematic uncertainties in extraction of the QE cross-section ratio.

| Source | Point-to-point (%) | Normalization (%) |
|---|---|---|
| Time-Dependent Instabilities | — | 1.0 |
| Target Thickness and Cuts | — | 1.42–1.58 |
| Acceptance Corrections | 1.2 (2.5,10) | — |
| Radiative Corrections | — | 0.5 |
| Coulomb Corrections | — | 0.2–1.0 |
| Bin-Centering Corrections | 0.5 | — |
| Kinematical Corrections | 0.3 | — |
| Total | 1.33 | 1.82–2.18 |

**Table V:** | **SRC Scaling Coefficients (This work).** Extracted SRC scaling coefficients and their uncertainties. Contributions to $a_2^n$ and $a_2^p$ can be obtained by scaling the $a_2$ values with $A/2N$ and $A/2Z$ respectively.

| Target | $a_2$ | Fit | Normalization | Acceptance Corrections | Bin Centering |
|---|---|---|---|---|---|
| | | | Contributions to the total uncertainty | | |
| $^{12}$C | 4.49 ± 0.17 | 0.08 | 0.08 | 0.09 | 0.07 |
| $^{27}$Al | 4.83 ± 0.18 | 0.10 | 0.09 | 0.10 | 0.07 |
| $^{56}$Fe | 4.80 ± 0.22 | 0.08 | 0.09 | 0.15 | 0.10 |
| $^{208}$Pb | 4.84 ± 0.20 | 0.09 | 0.11 | 0.11 | 0.08 |

**Table VI:** | **EMC Slopes (This work).** Extracted non isoscalar-corrected EMC Slopes ($dR_{\text{EMC}}/dx_B$) and the various contributions to their uncertainties. Contributions to $dR_{\text{EMC}}^n/dx_B$ and $dR_{\text{EMC}}^p/dx_B$ can be obtained by scaling the $dR_{\text{EMC}}/dx_B$ values with $A/2N$ and $A/2Z$ respectively.

| Target | $dR_{\text{EMC}}/dx_B$ | Fit | Normalization | Background | Acceptance | Bin Centering |
|---|---|---|---|---|---|---|
| | | | Contributions to the total uncertainty | | | |
| $^{12}$C | 0.340±0.022 | 0.019 | 0.006 | 0.004 | 0.002 | 0.007 |
| $^{27}$Al | 0.347±0.022 | 0.019 | 0.006 | 0.003 | 0.003 | 0.008 |
| $^{56}$Fe | 0.472±0.022 | 0.018 | 0.008 | 0.003 | 0.003 | 0.010 |
| $^{208}$Pb | 0.539±0.020 | 0.018 | 0.008 | 0.003 | 0.002 | 0.003 |

**Table VII:** | **Sensitivity of the EMC Slopes to cut variations.** Sensitivity of the extracted per-nucleon ($dR_{EMC}/dx_B$) non isoscalar-corrected EMC slopes from the current work to the kinematical selection cuts on $Q^2$ and $W$. As the kinematical cuts affect the $x_B$ acceptance (see Fig. 1), the extracted slopes are fit over a different range for each cut combination, as specified in the fit range column.

| Cuts | Fit Range | C/d | Al/d | Fe/d | Pb/d |
|---|---|---|---|---|---|
| $Q^2 > 1.5$ ; $W > 1.8$ | 0.25 − 0.56 | −0.340 ± 0.022 | −0.347 ± 0.022 | −0.472 ± 0.023 | −0.539 ± 0.020 |
| $Q^2 > 1.5$ ; $W > 2.0$ | 0.25 − 0.52 | −0.350 ± 0.026 | −0.366 ± 0.027 | −0.449 ± 0.027 | −0.538 ± 0.025 |
| $Q^2 > 1.75$ ; $W > 1.8$ | 0.28 − 0.55 | −0.344 ± 0.026 | −0.345 ± 0.027 | −0.477 ± 0.026 | −0.536 ± 0.024 |
| $Q^2 > 2.0$ ; $W > 1.8$ | 0.30 − 0.55 | −0.356 ± 0.028 | −0.301 ± 0.029 | −0.459 ± 0.028 | −0.505 ± 0.026 |
| $Q^2 > 2.5$ ; $W > 1.8$ | 0.38 − 0.55 | −0.310 ± 0.048 | −0.292 ± 0.051 | −0.468 ± 0.045 | −0.490 ± 0.045 |